\documentclass[10pt,twocolumn,letterpaper]{article}

\usepackage{iccv}
\usepackage{times}
\usepackage{epsfig}
\usepackage{graphicx}
\usepackage{amsmath}
\usepackage{amssymb}

\usepackage{booktabs}
\usepackage{xcolor}
\usepackage{amsmath, bm, subfigure, epstopdf, url, pifont, overpic, cases}
\usepackage{latexsym, amssymb, bbding, multirow, makecell,  diagbox, enumitem, caption}
\usepackage{array, enumitem, soul, algorithm, algpseudocode}

\def\comp{\ensuremath\mathop{\scalebox{.6}{$\circ$}}}

\usepackage{arydshln}

\usepackage[pagebackref=true,breaklinks=true,letterpaper=true,colorlinks,bookmarks=false]{hyperref}

\iccvfinalcopy %

\def\iccvPaperID{*} %

\begin{document}

\title{Hierarchical Conditional Flow: A Unified Framework for \\Image Super-Resolution and Image Rescaling}
\author{\hspace{-0.3cm}Jingyun Liang$^{1}$ ~ Andreas Lugmayr$^{1}$ ~ Kai Zhang$^{1,}$\thanks{Corresponding author.} ~ Martin Danelljan$^{1}$ ~ Luc Van Gool$^{1,2}$ ~ Radu Timofte$^{1}$\\
$^{1}$Computer Vision Lab, ETH Zurich, Switzerland\quad $^{2}$ KU Leuven, Belgium\\
{\tt\small \hspace{-0.6cm}\{jinliang, andreas.lugmayr, kai.zhang, martin.danelljan, vangool, timofter\}@vision.ee.ethz.ch}\\ 
{\tt\small }\url{https://github.com/JingyunLiang/HCFlow}
}

\maketitle

\begin{abstract}
Normalizing flows have recently demonstrated promising results for low-level vision tasks. For image super-resolution (SR), it learns to predict diverse photo-realistic high-resolution (HR) images from the low-resolution (LR) image rather than learning a deterministic mapping. For image rescaling, it achieves high accuracy by jointly modelling the downscaling and upscaling processes. While existing approaches employ specialized techniques for these two tasks, we set out to unify them in a single formulation. In this paper, we propose the hierarchical conditional flow (HCFlow) as a unified framework for image SR and image rescaling. More specifically, HCFlow learns a bijective mapping between HR and LR image pairs by modelling the distribution of the LR image and the rest high-frequency component simultaneously. In particular, the high-frequency component is conditional on the LR image in a hierarchical manner. To further enhance the performance, other losses such as perceptual loss and GAN loss are combined with the commonly used negative log-likelihood loss in training. Extensive experiments on general image SR, face image SR and image rescaling have demonstrated that the proposed HCFlow achieves state-of-the-art performance in terms of both quantitative metrics and visual quality.
\end{abstract}

\section{Introduction}
Normalizing flows~\cite{dinh2014nice, dinh2016realnvp, kingma2018glow, ho2019flow++, jaini2019polyflow, nielsen2020survae} are powerful deep generative probabilistic models that allow for efficient and exact likelihood calculation and sampling. They have been used in the generation of image~\cite{dinh2016realnvp, kingma2018glow}, blur kernel~\cite{liang2021fkp}, and audio~\cite{kim2018flowavenet} data. Recently, in the low-level vision community, normalizing flows have attracted much interest and have achieved promising progress for image super-resolution (SR)~\cite{lugmayr2020srflow} and image rescaling~\cite{xiao2020IRN}.

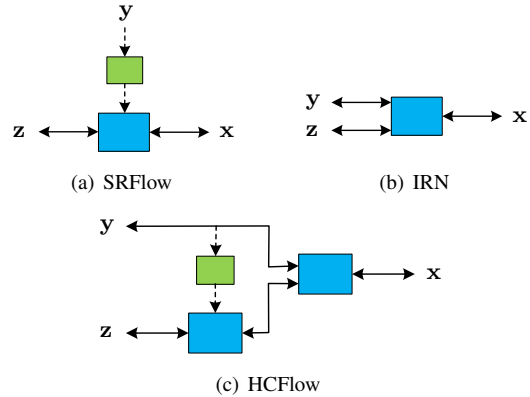
\begin{figure}[!tbp]
\captionsetup{font=small}%
\scriptsize
\begin{center}
\subfigcapskip=0.2cm
\subfigure[SRFlow]{
\begin{overpic}[width=0.13\textwidth]{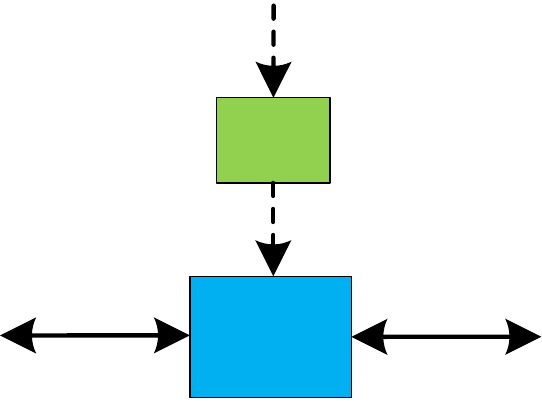}\label{fig:intro_srflow}
\put(106,9){\color{black}{\small $\mathbf{x}$}}
\put(47,80){\color{black}{\small $\mathbf{y}$}}
\put(-15,9){\color{black}{\small $\mathbf{z}$}}
\end{overpic}}\hspace{1.5cm}
\subfigure[IRN]{
\begin{overpic}[width=0.13\textwidth]{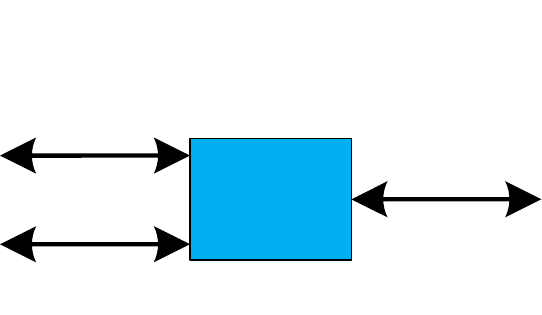}\label{fig:intro_irn}
\put(106,18){\color{black}{\small $\mathbf{x}$}}
\put(-15,28){\color{black}{\small $\mathbf{y}$}}
\put(-15,10){\color{black}{\small $\mathbf{z}$}}
\end{overpic}}\vspace{0.cm}
\subfigure[HCFlow]{
\begin{overpic}[width=0.22\textwidth]{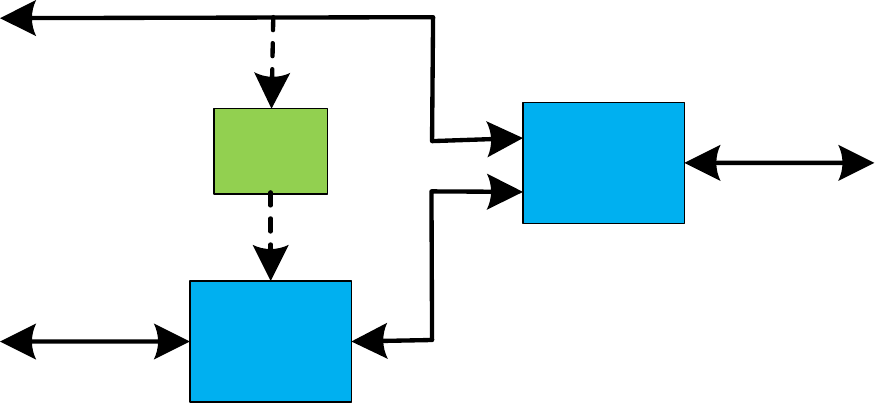}\label{fig:intro_mcflow}
\put(104,26){\color{black}{\small $\mathbf{x}$}}
\put(-9,43){\color{black}{\small $\mathbf{y}$}}
\put(-9,6){\color{black}{\small $\mathbf{z}$}}
\end{overpic}}
\end{center}
\vspace{-0.7cm}
\caption{The comparison between SRFlow~\cite{lugmayr2020srflow}, IRN~\cite{xiao2020IRN} and the proposed HCFlow. $\mathbf{x}$, $\mathbf{y}$ and $\mathbf{z}$ denote HR image, LR image and the latent variable, respectively. Blue boxes are invertible neural networks, while green ones are non-invertible models (\eg, CNN). Solid bi-directional arrows denote bijective mappings, while dashed arrows represent conditional relations.}
\label{fig:intro}
\vspace{-0.2cm}
\end{figure}

SRFlow~\cite{lugmayr2020srflow} is a seminal flow-based model for image SR. Unlike previous CNN-based models that learn a deterministic mapping from the low-resolution (LR) image to the high-resolution (HR) image, SRFlow learns the distribution of HR images and is able to generate diverse photo-realistic HR images. However, as shown in Fig.~\ref{fig:intro_srflow}, it treats the LR image as an external conditional prior and thus is not fully invertible between HR and LR image pairs, making it hard to be used for image rescaling. Another work IRN~\cite{xiao2020IRN} employs an invertible neural network to learn downscaling and upscaling for image rescaling. Since the model is bijective, it can recover the input HR image with high accuracy after downscaling. Nevertheless, as shown in Fig.~\ref{fig:intro_irn}, it assumes the high-frequency and low-frequency components of the image are independent to each other and thus lacks the ability to exploit their dependency for image SR.

\subfigcapskip=0.1cm

In this paper, we propose a hierarchical conditional flow (HCFlow) as a unified framework for both image SR and rescaling. As shown in Fig.~\ref{fig:intro_mcflow}, HCFlow is an invertible flow-based model for modelling the HR-LR relationship, in which the high-frequency component is hierarchically conditional on the low-frequency component of the image. More specifically, in the forward propagation, HCFlow learns to decompose the input HR image into the LR image and a latent variable. In the inverse propagation, it generates HR images based on the LR input and random samples of the latent variable. The modelling of the latent variable (high-frequency component) is conditional on the generated LR image (low-frequency component) in a hierarchical manner.

When trained for image SR, HCFlow is optimized by minimizing the negative log-likelihood loss on the basis of tractable Jacobian determinant computation. To further improve visual quality, we integrate a pixel loss, perceptual loss, and GAN loss in the inverse propagation to constrain the learned HR space. Moreover, HCFlow can be used for the image rescaling task. It can decompose the HR image to a visually-pleasing LR image and a latent variable that follows a simple distribution. In this case, HCFlow is trained as an encoder-decoder framework, in which the forward and inverse processes are jointly optimized. As HCFlow is bijective, it can recover the HR image faithfully by sampling from the latent space given the generated LR image.

Our contributions can be summarized as follows:
\begin{itemize}
  \vspace{-0.2cm}
  \item[1)] We propose a unified framework for image SR and image rescaling. It learns to model the LR image and the residual high-frequency component simultaneously. The high-frequency component is hierarchically conditional on the generated LR image.
  \vspace{-0.2cm}
  \item[2)] We propose additional losses to train normalizing flows, including pixel, perceptual, and GAN losses, which effectively enhances the HR image quality. 
  \vspace{-0.2cm}
  \item[3)] We perform extensive experiments on three tasks: general image SR, face image SR and image rescaling. HCFlow achieves state-of-the-art results on all tasks in terms of both quantitative metrics and visual quality.
\end{itemize}

\section{Related Work}
In this section, we will briefly review image SR and image rescaling with a particular focus on two highly related flow-based methods, \ie, SRFlow~\cite{lugmayr2020srflow} and IRN~\cite{xiao2020IRN}.

\label{sec:related_work}
\subsection{Image SR}
Image SR aims to reconstruct the HR image given the LR image. Since the pioneer work SRCNN~\cite{dong2014srcnn}, many CNN-based models have been proposed in recent years~\cite{dong2014srcnn, kim2016vdsr, ledig2017SRGAN, wang2018esrgan, zhang2018rcan, zhang2018RDN, liu2020RFANet, zhang2020usrnet, liang21swinir, kai2021bsrgan, liang21manet, zhang2021DPIR}. Most of them focus on delicate feature extraction module design and generate over-smoothed images when trained with the pixel loss. To remedy this, the perceptual loss~\cite{johnson2016perceptual, wang2018esrgan} and GAN loss~\cite{goodfellow2014generative, ledig2017SRGAN, wang2018esrgan, zhang2019ranksrgan} are introduced to improve the perceptual quality. Despite of above progresses, they usually learn a deterministic mapping between the LR image and HR image, which is unnatural for image SR since one LR image may correspond to multiple HR images. 

\vspace{0.1cm}
\noindent \textbf{SRFlow}~\cite{lugmayr2020srflow}\textbf{.}
Normalizing flows~\cite{dinh2014nice, dinh2016realnvp, kingma2018glow, ho2019flow++, jaini2019polyflow, nielsen2020survae, wolf2021deflow} provide a new possible solution for image SR. SRFlow designs a conditional flow to model the distribution of HR images, conditional on LR images. It can generate diverse photo-realistic images by sampling different latent variables. Our proposed HCFlow differs from SRFlow in two main aspects: First, SRFlow uses the LR image as an external conditional prior and maps the HR distribution to a simple latent distribution. Therefore, it cannot generate LR image and thus is not applicable for image rescaling. In contrast, HCFlow models the LR image and treats it as part of the latent space. Second, SRFlow basically follows the flow framework proposed in~\cite{dinh2016realnvp}, while HCFlow proposes a new framework with hierarchical conditional mechanism.

\subsection{Image Rescaling}
Image rescaling aims to downscale the HR image to a visually meaningful LR image, and then recover the HR image plausibly. Different from image SR that works on a given LR image space, image rescaling tries to maintain as much information from the HR image as possible for a better subsequent reconstruction, for the purpose of reducing the storage and bandwidth cost. In other words, it can define its own LR image space which is expected to be more informative than that by simple downscaling such as bicubic downscaling. In general, in image rescaling, the downscaling and upscaling processes are jointly modelled by an encoder-decoder framework~\cite{kim2018task, li2018learning, sun2020learned}, so that the downscaling model is optimized for the later upscaling operation.

\vspace{0.1cm}
\noindent
\textbf{IRN}~\cite{xiao2020IRN}\textbf{.}
Recently, IRN proposes to use a bijective invertible neural network to model the downscaling and upscaling processes. High-frequency component is well-captured and transformed to a structured latent space in training. In testing, the HR image can be recovered by inputting the generated LR image and a randomly sampled latent variable. In particular, IRN assumes the LR image and high-frequency component is independent to each other. These two components are divided apart and learned separately. By contrast, HCFlow assumes the removed high-frequency component is dependent on the LR image and thus employs a hierarchical conditional framework to model the LR image and the conditional distribution of the high-frequency component. Besides, although IRN designs a bijective mapping between HR and LR image pairs, it can only be trained by Monte Carlo simulation rather than maximum likelihood estimation (MLE). HCFlow can be trained in the same way for image rescaling, but it further models the LR image distribution and allows for tractable Jacobian determinant computation, making it possible for probabilistic modelling of HR and LR images when trained by MLE.

\begin{figure*}[!tbp]
\captionsetup{font=small}%
\scriptsize
\vspace{-0.5cm}
\begin{center}
\subfigcapskip=0.15cm
\subfigure[Forward propagation]{
\begin{overpic}[width=0.35\textwidth]{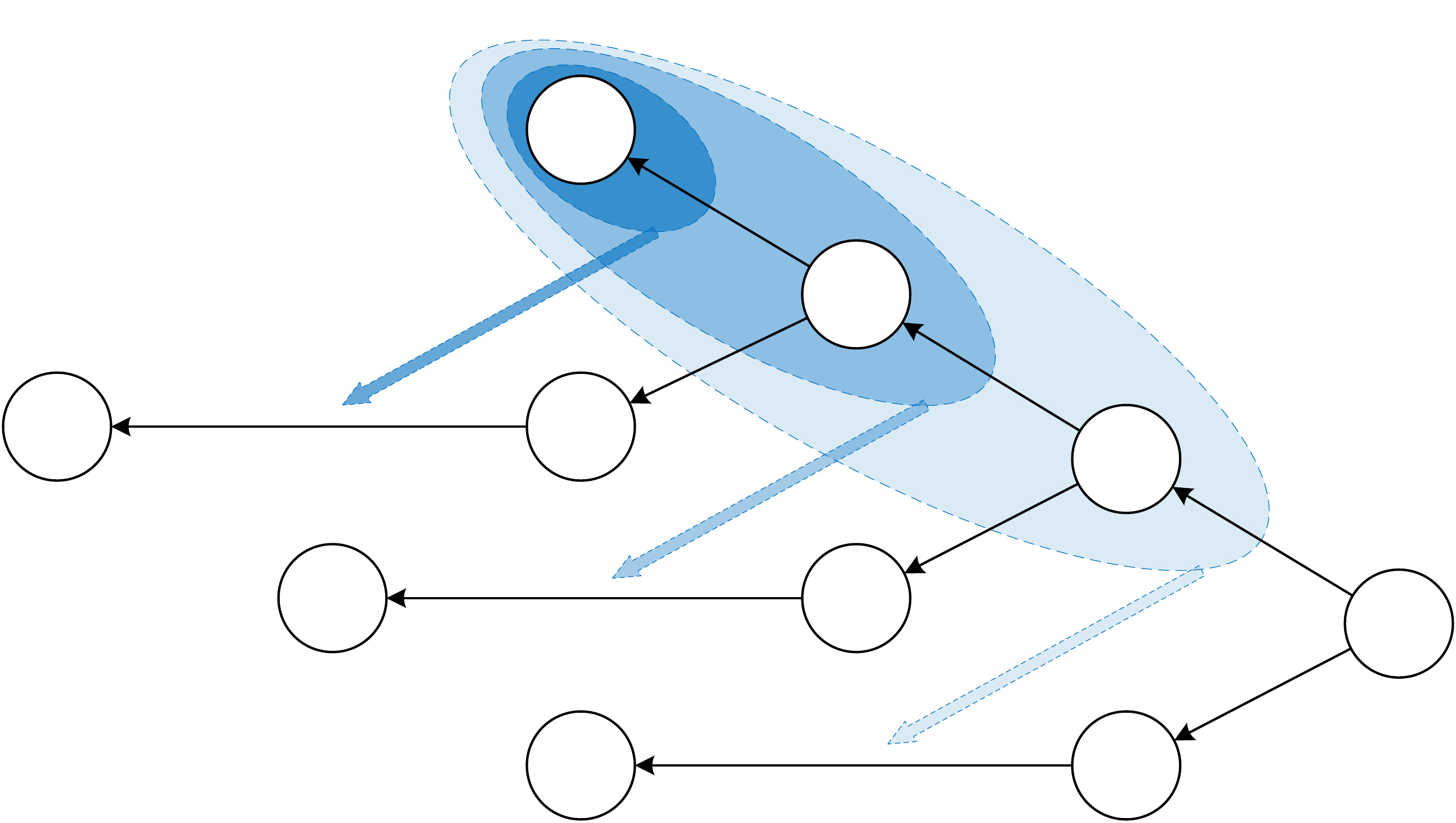}
\put(94.8,12.8){\color{black}{\small $\mathbf{x}$}}
\put(75.5,24.5){\color{black}{\small $\mathbf{y}_1$}}
\put(75.5,3){\color{black}{\small $\mathbf{a}_1$}}
\put(38,3){\color{black}{\small $\mathbf{z}_1$}}
\put(57,35.5){\color{black}{\small $\mathbf{y}_2$}}
\put(57,14.5){\color{black}{\small $\mathbf{a}_2$}}
\put(21,14.5){\color{black}{\small $\mathbf{z}_2$}}
\put(38,47){\color{black}{\small $\mathbf{y}_3$}}
\put(38,26){\color{black}{\small $\mathbf{a}_3$}}
\put(2.2,26){\color{black}{\small $\mathbf{z}_3$}}

\put(94.8,18.8){\color{black}{\small $1$}}
\put(75.8,30.5){\color{black}{\small $2$}}
\put(75.8,9){\color{black}{\small $9$}}
\put(37.3,9){\color{black}{\small $10$}}
\put(57.3,41.5){\color{black}{\small $3$}}
\put(57.3,20.5){\color{black}{\small $7$}}
\put(21.2,20.5){\color{black}{\small $8$}}
\put(38.5,54){\color{black}{\small $4$}}
\put(38.5,32){\color{black}{\small $5$}}
\put(2.4,32){\color{black}{\small $6$}}
\end{overpic}}
\hspace{1.5cm}
\subfigure[Inverse propagation]{
\begin{overpic}[width=0.35\textwidth]{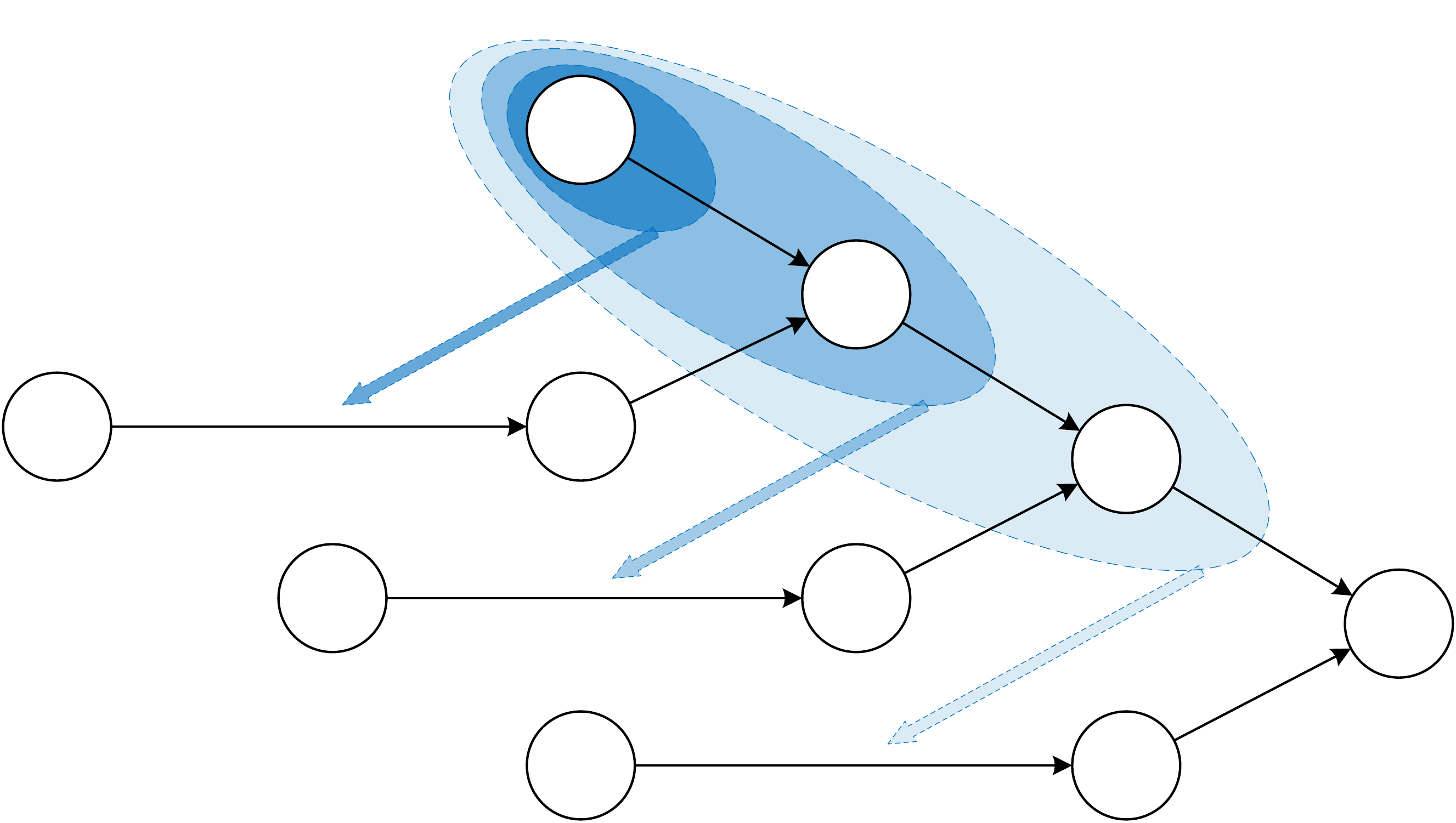}
\put(94.8,12.8){\color{black}{\small $\mathbf{x}$}}
\put(75.5,24.5){\color{black}{\small $\mathbf{y}_1$}}
\put(75.5,3){\color{black}{\small $\mathbf{a}_1$}}
\put(38,3){\color{black}{\small $\mathbf{z}_1$}}
\put(57,35.5){\color{black}{\small $\mathbf{y}_2$}}
\put(57,14.5){\color{black}{\small $\mathbf{a}_2$}}
\put(21,14.5){\color{black}{\small $\mathbf{z}_2$}}
\put(38,47){\color{black}{\small $\mathbf{y}_3$}}
\put(38,26){\color{black}{\small $\mathbf{a}_3$}}
\put(2.2,26){\color{black}{\small $\mathbf{z}_3$}}

\put(93.8,18.8){\color{black}{\small $10$}}
\put(75.8,30.5){\color{black}{\small $7$}}
\put(75.8,9){\color{black}{\small $9$}}
\put(38.5,9){\color{black}{\small $8$}}
\put(57.3,41.5){\color{black}{\small $4$}}
\put(57.3,20.5){\color{black}{\small $6$}}
\put(21.2,20.5){\color{black}{\small $5$}}
\put(38.5,54){\color{black}{\small $1$}}
\put(38.5,32){\color{black}{\small $3$}}
\put(2.4,32){\color{black}{\small $2$}}
\end{overpic}}
\end{center}\vspace{-0.3cm}
\caption{Schematic computational graphs of the hierarchical conditional flow (HCFlow) with 3 flow levels. On level $l$, $\mathbf{y}_{l-1}$ (note that $\mathbf{y}_0=\mathbf{x}$) is decomposed to low-frequency component $\mathbf{y}_{l}$ and high-frequency component $\mathbf{a}_{l}$. The transformation between $\mathbf{a}_{l}$ and $\mathbf{z}_{l}$ is conditional on $[\mathbf{y}_{L},\mathbf{y}_{L-1},...,\mathbf{y}_{l}]$, as indicated by the blue arrows. The computation orders in forward and inverse propagation are shown on the top of each node.}
\label{fig:flow_tree}
\end{figure*}
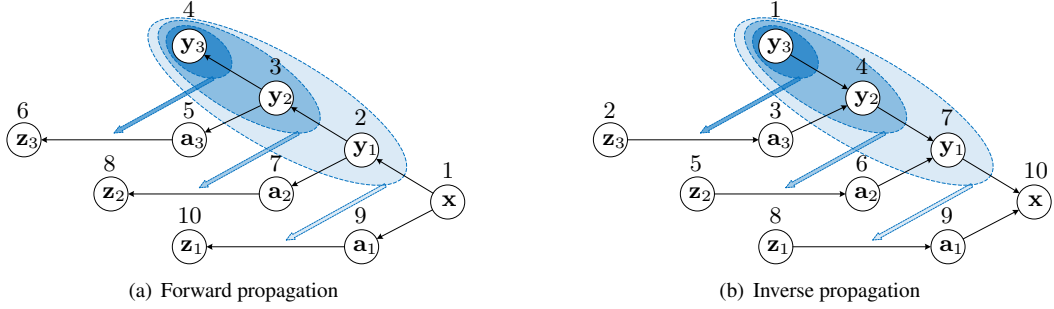
\subfigcapskip=0.1cm

\section{Methodology}
\subsection{Preliminaries}
Flow-based models~\cite{dinh2014nice, dinh2016realnvp, kingma2016iaf, papamakarios2017maf, huang2018naf, kingma2018glow, ardizzone2018analyzingIRN, jaini2019polyflow, ho2019flow++, nielsen2020survae, liang2021fkp} aim to learn a bijective mapping between the target space and the latent space.
For a high-dimensional random variable (\eg, an image) $\mathbf{x}$ with distribution $\mathbf{x}\sim p(\mathbf{x})$ and a latent variable $\mathbf{z}$ with simple tractable distribution $\mathbf{z}\sim p(\mathbf{z})$ (\eg, multivariate Gaussian distribution), flow models generally use an invertible neural network $f_{\bm\theta}$ to transform $\mathbf{x}$ to $\mathbf{z}$: $\mathbf{z}=f_{\bm\theta}(\mathbf{x})$.
Conversely, $\mathbf{x}$ can be recovered from $\mathbf{z}$ by the inverse mapping $\mathbf{x}=f_{\bm\theta}^{-1}(\mathbf{z})$. 

Generally, $f_{\bm\theta}$ is composed of a series of invertible transformations: $f_{\bm{\theta}} = f_{\bm{\theta}}^1 \comp f_{\bm{\theta}}^2 \comp \cdots \comp f_{\bm{\theta}}^K$. The intermediate variables are defined as $\mathbf{h}^{k}=f_{\bm{\theta}}^k(\mathbf{h}^{k-1})$ for $k\in\{1,...,K\}$. The input $\mathbf{h}^{0}$ and output $\mathbf{h}^{N}$ of $f_{\bm{\theta}}$ are $\mathbf{x}$ and $\mathbf{z}$, respectively. Concretely, $f_{\bm{\theta}}^k$ are flow layers such as squeeze layer, batch normalization layer, affine coupling layer, \etc.

According to the change of variable formula and the chain rule, for a sample $\mathbf{x}$, the log probability $\log(\mathbf{x})$ can be calculated as
\begin{equation}\hspace{-0.2cm}
\begin{aligned}
\log p (\mathbf{x})
&= \log p (f_{\bm{\theta}}(\mathbf{x})) 
+\sum_{k=1}^{K} \log
\left\lvert 
\det\frac{\partial f_{\bm{\theta}}^k (\mathbf{h}^{k-1})}{\partial \mathbf{h}^{k-1}} 
\right\rvert,\hspace{-0.5cm}
\end{aligned}
\label{eq:nll}
\end{equation}
where $\log
\left\lvert 
\det\frac{\partial f_{\bm{\theta}}^k (\mathbf{h}^{k-1})}{\partial \mathbf{h}^{k-1}} 
\right\rvert$ is the logarithm of the absolute value of the determinant of the Jacobian of $f_{\bm{\theta}}^k$ at $\mathbf{h}^{k-1}$. The flow model can thereby be optimized by minimizing the negative log-likelihood loss.

\begin{figure*}[!tbp]
\captionsetup{font=small}%
\scriptsize
\begin{center}
\hspace{-0.1cm}
\begin{overpic}[width=17.3cm]{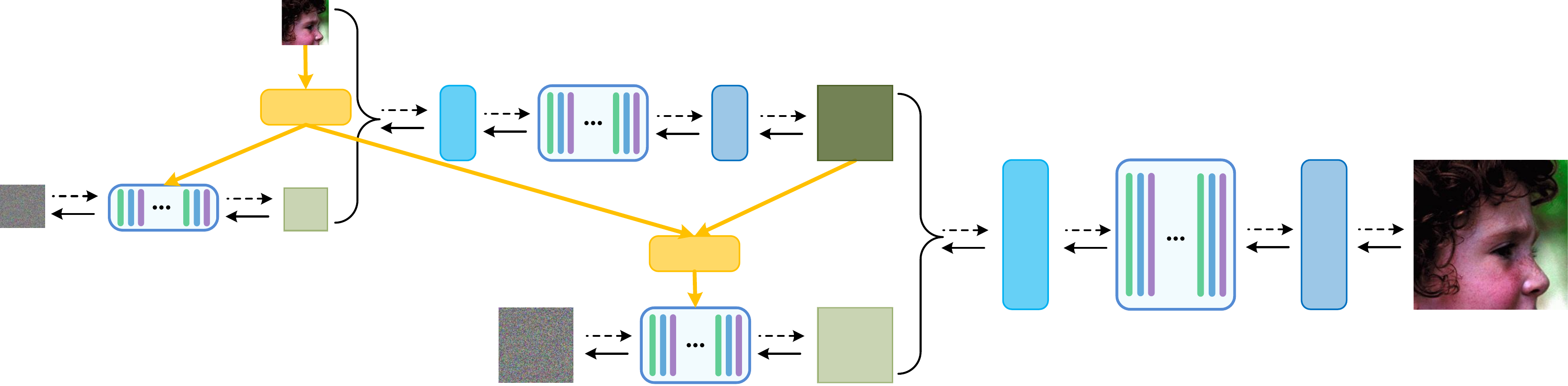}
\put(18.5,25.8){\color{black}{\small $\mathbf{y}_2$}}
\put(0.7,8){\color{black}{\small $\mathbf{z}_2$}}
\put(18.7,8){\color{black}{\small $\mathbf{a}_2$}}
\put(53.5,20.5){\color{black}{\small $\mathbf{y}_1$}}
\put(33.5,-2){\color{black}{\small $\mathbf{z}_1$}}
\put(53.9,-2){\color{black}{\small $\mathbf{a}_1$}}
\put(94,15.5){\color{black}{\small $\mathbf{x}$}}
\put(83.3,13.){\color{black}{ \fontsize{8}{5}\selectfont \rotatebox{270}{ {\makecell{{Squeeze}}}}}}
\put(64.2,11.8){\color{black}{ \fontsize{8}{5}\selectfont \rotatebox{270}{ {\makecell{{Split}}}}}}
\put(45.6,19.1){\color{black}{ \fontsize{6}{4}\selectfont \rotatebox{270}{ {\makecell{{Squeeze}}}}}}
\put(28.2,18.2){\color{black}{ \fontsize{6}{4}\selectfont \rotatebox{270}{ {\makecell{{Split}}}}}}
\put(16.8,17.3){\color{black}{ \fontsize{6}{4}\selectfont \rotatebox{0}{ {\makecell{Feature\\Extractor}}}}}
\put(41.5,8){\color{black}{ \fontsize{6}{4}\selectfont \rotatebox{0}{ {\makecell{Feature\\Extractor}}}}}

\end{overpic}
\end{center}%
\caption{The architecture of the hierarchical conditional flow (HCFlow) with 2 flow levels. For a HR image $\mathbf{x}$, we first squeeze, transform and split it to low-frequency component $\mathbf{y}_1$ and high-frequency component $\mathbf{a}_1$. Similarly, $\mathbf{y}_1$ is decomposed to $\mathbf{y}_2$ (\ie, the LR image in this case) and $\mathbf{a}_2$ in the next level. $\mathbf{a}_1$ and $\mathbf{a}_2$ are transformed to latent variables $\mathbf{z}_1$ and $\mathbf{z}_2$, conditional on $\phi_1([\phi_2(\mathbf{y}_2),\mathbf{y}_1])$ and $\phi_2(\mathbf{y}_2)$ (note that $\phi_1$ and $\phi_2$ are feature extractors, \eg, CNN) respectively, in a hierarchical manner. The model is trained by negative log-likelihood loss, and can be further enhanced by pixel loss, perceptual loss and GAN loss.}
\label{fig:flow_main}
\end{figure*}
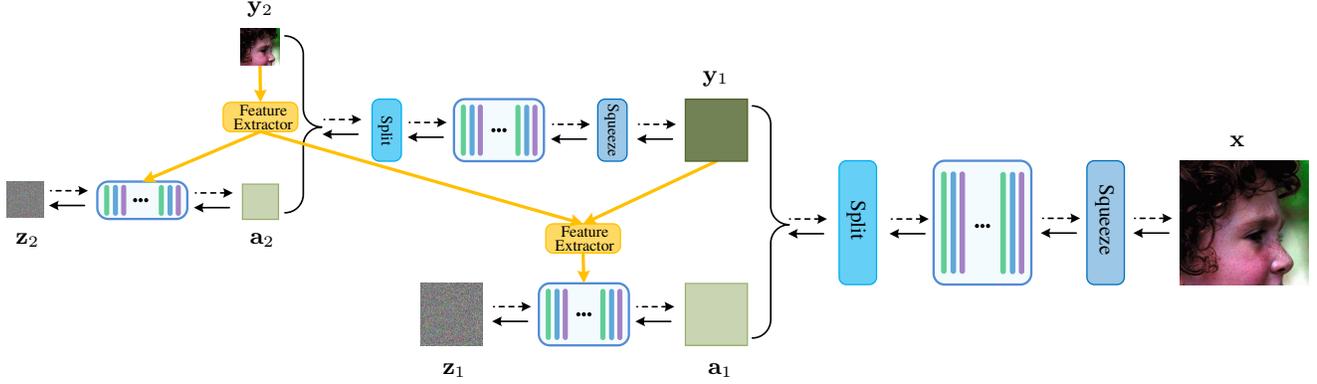

\subsection{Model Specification}
Both image SR and image rescaling try to reconstruct the HR image $\mathbf{x}$ given a LR image. Since the image degradation process (or image downscaling) is the inverse of image super-resolution (or image upscaling), we can model these two processes with an invertible bijective transformation: $\mathbf{x}\leftrightarrow [\mathbf{y},\mathbf{a}]$, where $\mathbf{y}$ and $\mathbf{a}$ are the generated LR image and the rest high-frequency component, respectively. As modelling the probability of natural images is a non-trivial task, it is reasonable to design a flow model conditional on the ground-truth LR image $\mathbf{y}^*$ as,
\begin{equation}
p(\mathbf{x}|\mathbf{y}^*)\leftrightarrow p(\mathbf{y},\mathbf{a}|\mathbf{y}^*)=p(\mathbf{y}|\mathbf{y}^*)p(\mathbf{a}|\mathbf{y},\mathbf{y}^*).
\label{model1}
\end{equation} 

Ideally, we hope the model can generate exactly the same LR image as the ground-truth LR image. This can be formulated as a Dirac delta function $\delta(\mathbf{y}-\mathbf{y}^*)$ and further approximated by a multivariate Gaussian distribution as,
\begin{equation}
\begin{aligned}
p(\mathbf{y}|\mathbf{y}^*)p(\mathbf{a}|\mathbf{y},\mathbf{y}^*)
&=\delta(\mathbf{y}-\mathbf{y}^*)p(\mathbf{a}|\mathbf{y})\\
&=\lim\limits_{\mathbf\Sigma\to \mathbf{0}}\mathcal{N}(\mathbf{y}|\mathbf{y}^*,\mathbf\Sigma)p(\mathbf{a}|\mathbf{y}),
\end{aligned}
\label{model2}
\end{equation}
where $\mathbf\Sigma$ is a diagonal covariance matrix with all diagonal elements close to zero. Note that $\textbf{y}$ is nearly equal to $\textbf{y}^*$ in this case. By further mapping $p(\mathbf{a}|\mathbf{y})$ to a standard multivariate Gaussian distribution $p(\mathbf{z})=\mathcal{N}(\mathbf{z}|\mathbf{0},\mathbf{I})$, the flow model is defined as,
\begin{equation}
\begin{aligned}
p(\mathbf{x}|\mathbf{y}^*)
&\leftrightarrow \lim\limits_{\mathbf\Sigma\to \mathbf{0}}\mathcal{N}(\mathbf{y}|\mathbf{y}^*,\mathbf\Sigma)\mathcal{N}(\mathbf{z}|\mathbf{0},\mathbf{I}).
\end{aligned}
\label{model3}
\end{equation} 

As we can see, part of the latent space is constrained to be the LR image space. In particular, decomposed high-frequency component $\mathbf{a}$ is conditional on another decomposed component $\mathbf{y}$. Once trained, following the forward direction, HCFlow can decompose the HR image $\mathbf{x}$ into LR image $\mathbf{y}$ and latent variable $\mathbf{z}$ that follows a simple distribution. Following the inverse direction, HCFlow can generate $\mathbf{x}$ given the LR image input $\mathbf{y}^*$ and a random sample $\mathbf{z}$ from the latent distribution, as it is an invertible bijective model.

Note that this model regards $\mathbf{y}^*$ as an input or output, rather than as an external conditional prior. Therefore, it is not explicitly conditional on $\mathbf{y}^*$ and is fully invertible between HR and LR image pairs. Besides, by approximating the the distribution of $\mathbf{y}$ with a multi-variate Gaussian distribution, it allows for tractable Jacobian determinant computation, so that the model can be optimized by maximum likelihood estimation (MLE).

\subsection{Model Architecture}
The multi-scale architecture proposed in RealNVP~\cite{dinh2016realnvp} is a popular normalizing flow architecture~\cite{lugmayr2020srflow, kingma2018glow, ho2019flow++}. It consists of $L$ levels and at the end of each level, half of the dimensions are factored out. Generally, the factored out dimensions are directly Gaussianized for the computation of negative log-likelihood loss, lacking sufficient modelling of these dimensions. Therefore, based on the multi-scale architecture, we take a further step to model factored out dimensions conditional on the reserved dimensions.

As illustrated in Fig.~\ref{fig:flow_tree}, at each level $l$, $\mathbf{y}_{l-1}$ is decomposed to low-frequency component $\mathbf{y}_l$ and high-frequency component $\mathbf{a}_l$. Then, $\mathbf{a}_l$ is modelled by an additional flow that is conditional on the concatenation of tensors $\mathbf{y}_L, \mathbf{y}_{L-1},..., \mathbf{y}_l$ from multiple flow levels. By this design, the reconstruction of high-frequency component is hierarchically conditional on frequencies reconstructed from all previous levels. In forward propagation, similar to the depth-first traversal of a binary tree, we first compute $\mathbf{y}_1$, $\mathbf{y}_2$, ..., $\mathbf{y}_L$ in order. Then, we model the factored out dimensions in a reverse order: $\mathbf{a}_L$, $\mathbf{a}_{L-1}$, ..., $\mathbf{a}_1$. In inverse propagation, we compute $\mathbf{y}_l$ and $\mathbf{a}_l$ level by level, from level $L$ to level $1$. Note that the determinant of Jacobian of the whole flow can still be efficiently computed, since the conditional relations between $\mathbf{a}_1, \mathbf{a}_{2},..., \mathbf{a}_L$ and $\mathbf{y}_1, \mathbf{y}_{2},...,  \mathbf{y}_L$ can be represented as an upper triangle block matrix.

The detailed architecture of HCFlow is shown in Fig.~\ref{fig:flow_main}. For each level, the first layer is the squeeze layer, which transforms the $H\times W\times C$ input to a $\frac{H}{2}\times\frac{W}{2}\times 4C$ tensor by trading spatial size for number of channels. Then, $K$ flow-steps are used for transforming the tensor and decomposing it into different components. More specifically, each flow-step consists of a sequence of three layers: Actnorm layer, invertible $1\times 1$ convolution layer and affine coupling layer~\cite{dinh2016realnvp, kingma2018glow}. After that, the split layer is used to evenly split the tensor into two tensors $\mathbf{y}_l\in\mathcal{R}^{\frac{H}{2}\times\frac{W}{2}\times 2C}$ and $\mathbf{a}_l\in\mathcal{R}^{\frac{H}{2}\times\frac{W}{2}\times 2C}$ along the channel dimension. Note that, for the last level, we only keep 3 channels for $\mathbf{y}_l$ to make it fit the RGB space of the LR image. Next, $\mathbf{y}_l$ is fed to the next level, while $\mathbf{a}_l$ is input into an additional flow.

In the $l$-th additional flow, $\mathbf{a}_l$ is transformed to the latent variable $\mathbf{z}_l$ by $P$ flow-steps. Different from above flow-steps, we use conditional affine coupling layer~\cite{ardizzone2019guided, winkler2019learning} rather than ordinary affine coupling layer to obtain a conditional flow. In particular, we first upscale the conditional feature $\mathbf{c}_{l+1}$ from level $l+1$ by $\times 2$ nearest neighbor interpolation, and concatenate it with $\mathbf{y}_l$. Then, we use a feature extractor $\phi_l$ to extract image features, which act as the conditional feature $\mathbf{c}_{l}$ for level $l$.
Note that the feature extractor only provides scale and shift for an affine coupling during both forward and inverse propagation. Hence the constrains on being invertible and having a tractable Jacobian do not hold for this part.
More formally, the hierarchical conditional mechanism of HCFlow is formulated as follows,
\begin{equation}\hspace{-0.4cm}
\mathbf{c}_l=
\begin{cases}
\phi_l(\mathbf{y}_l)& l=L\\
\phi_l([\mathbf{c}_{L},\mathbf{c}_{L-1},...,\mathbf{c}_{l+1},\mathbf{y}_l])& l=L-1,...,1
\end{cases},
\end{equation}
where conditional features of different levels are computed in a reverse order, from $\mathbf{c}_L$ to $\mathbf{c}_1$.

Particularly, for the last level, we directly model $\mathbf{y}_L$ by a Dirac delta function $\delta(\mathbf{y}-\mathbf{y}^*)$ instead of transforming it to another latent variable. This constrains part of the latent space to be the LR image space and implicitly makes the model be conditional on $\mathbf{y}^*$.

\subsection{Training Objectives}
\paragraph{Image SR.} When HCFlow is used for image SR, it can be trained by minimizing the negative log-likelihood loss,
\begin{equation}
\begin{aligned}
\mathcal{L}_{nll}=-\log p (\mathbf{x}),
\end{aligned}
\end{equation}
which is unsupervised and converges stably. However, in practice, this loss converges slowly and does not provide strong supervision for image SR. To achieve better HR image PSNR, we can add $\mathcal{L}_1$ pixel loss on the generated SR image in inverse propagation, leading to a loss function as follows,
\begin{equation}
\begin{aligned}
\mathcal{L}=\lambda_1 \mathcal{L}_{nll} (\mathbf{x})
+\lambda_2 \mathcal{L}_{pixel}(\mathbf{x},\mathbf{x}_{\tau=0}),
\end{aligned}
\end{equation}
where $\mathbf{x}$ is the ground-truth HR image and $\mathbf{x}_{\tau=0}$ is the generated SR image by inputting the ground-truth LR image $\mathbf{y}^*$ and sampling the latent variable $\mathbf{z}$ with temperature $\tau=0$. The added pixel loss can help the flow to learn the SR manifold centered around the PSNR-oriented SR image. Furthermore, we can add perceptual loss~\cite{johnson2016perceptual} and GAN loss~\cite{goodfellow2014generative} on the generated SR image to improve the visual quality. This is formulated as,
\begin{equation}
\begin{aligned}
\mathcal{L}=&\lambda_1 \mathcal{L}_{nll} (\mathbf{x})
+\lambda_2 \mathcal{L}_{pixel}(\mathbf{x},\mathbf{x}_{\tau=0})\\
&+\lambda_3 \mathcal{L}_{percep}(\mathbf{x},\mathbf{x}_{\tau=\tau_0})
+\lambda_4 \mathcal{L}_{gan}(\mathbf{x},\mathbf{x}_{\tau=\tau_0}),
\end{aligned}
\end{equation}
where $\mathbf{x}_{\tau=\tau_0}$ is the generated SR image by inputting $\mathbf{y}^*$ and sampling $\mathbf{z}$ with $\tau=\tau_0$. Note that unlike the pixel loss that uses $\tau=0$, $\tau_0$ is set to 0.8 or 0.9 to preserve the diversity of HR images.

\vspace{-0.4cm}
\paragraph{Image rescaling.}
Different from image SR, image rescaling aims to recover exactly the same HR image. Following ~\cite{xiao2020IRN}, we regard the invertible HCFlow as an encoder-decoder framework, in which the forward and inverse processes correspond to the encoding and decoding stages, respectively. The loss is as follows,
\begin{equation}
\begin{aligned}
\mathcal{L}=&\lambda_1 \mathcal{L}_{pixel\_hr}(\mathbf{x},\mathbf{x}_{\tau=1})
+ \lambda_2 \mathcal{L}_{pixel\_lr}(\mathbf{y^*},\mathbf{y}) \\
&+\lambda_3 \mathcal{L}_{latent} (\mathbf{z}),
\end{aligned}
\end{equation}
where $\mathcal{L}_{pixel\_hr}$ is the $\mathcal{L}_1$ pixel loss to ensure that, after downscaling and upscaling, the reconstructed image $\mathbf{x}_{\tau=1}$ is close to the input $\mathbf{x}$. Note that this loss would dramatically decrease the diversity of generated images. Besides, $\mathcal{L}_{pixel\_lr}$ is the $\mathcal{L}_2$ pixel loss on the LR image, which guides $\mathbf{y}$ to be close to the bicubic LR image $\mathbf{y^*}$, so as to generate visually-pleasing LR images in downscaling. The last term $\mathcal{L}_{latent} (\mathbf{z})$ is the $\mathcal{L}_2$ regularization on the latent variable $\mathbf{z}$.

\section{Experiments}
\subsection{Experimental Setup}
We conduct experiments on general image SR, face image SR and image rescaling to show the effectiveness of HCFlow. For image SR experiments, we train the model by three loss combinations: $\mathcal{L}_{nll}$, $\mathcal{L}_{nll}+\mathcal{L}_{pixel}$ and $\mathcal{L}_{nll}+\mathcal{L}_{pixel}+\mathcal{L}_{percep}\&\mathcal{L}_{gan}$. The corresponding learned models are denoted as \textbf{HCFlow}, \textbf{HCFlow+} and \textbf{HCFlow++}, respectively.

\vspace{-0.4cm}
\paragraph{Image SR.}
For general image SR ($\times 4$), we set $L, K, P$ to 2, 13 and 13, respectively. Two 13-block RRDB networks~\cite{wang2018esrgan} are used as feature extractors. More details on the architecture are provided in the supplementary. The model is trained on the training set of DIV2K~\cite{DIV2K} and Flickr2K~\cite{Flickr2K} with random flips. The crop patch size and mini-batch size are set to $160\times 160$ and $16$, respectively. Adam optimizer~\cite{kingma2014adam} with $\beta_1=0.9$ and $\beta_2=0.99$ is used for optimization. For HCFLow (with only $\mathcal{L}_{nll}$), the learning rate is $2.5\times 10^{-4}$ and reduced by half at $50\%$, $75\%$, $90\%$ and $95\%$ of $300k$ iterations. We fine-tune HCFLow+ (with $\mathcal{L}_{nll}+\mathcal{L}_{pixel}$) for $50k$ iterations from the pretrained HCFlow. The weight of $\mathcal{L}_{nll}$ and $\mathcal{L}_{pixel}$ are $\lambda_1=2\times 10^{-3}$ and $\lambda_2=1$, respectively. It is worth pointing out that we can achieve even higher PSNR (about 0.2dB) if we train HCFlow+ from scratch. Similarly, we can fine-tune HCFlow++ by further adding $\mathcal{L}_{percep}$ and $\mathcal{L}_{gan}$. The loss weighting parameters are $\lambda_1=2\times 10^{-3}$, $\lambda_2=1$, $\lambda_3=5\times 10^{-2}$ and $\lambda_4=5\times 10^{-1}$.

For face image SR ($\times 8$), $L, K, P$ are set to 3, 13 and 13, respectively. Three 8-block RRDB networks are used as feature extractors. We train the model on the CelebA training set~\cite{liu2015celeba} and test it using first 5,000 images from the testing set. Following \cite{kingma2018glow, lugmayr2020srflow}, we crop and resize the HR images to the resolution of $160\times 160$, and flip them randomly for data augmentation. Other training details are the same as general image SR.

\begin{table*}[!thbp]
\captionsetup{font=small}
\footnotesize
\center
\begin{center}
\caption[Caption for LOF]{Ablation study on latent space and conditional priors for general image SR ($\times 4$). Results are tested on DIV2K~\cite{DIV2K} validation set.}\vspace{-0.2cm}
\label{tab:ablation_latent}
\begin{tabular}{c|c|c|c|p{1.2cm}<{\centering}|p{1.2cm}<{\centering}|p{1.6cm}<{\centering}|p{1.6cm}<{\centering}|p{1.4cm}<{\centering}}
\hline
 Case & Latent Space &  \makecell{Conditional Prior\\($l=2$)} & \makecell{Conditional Prior\\($l=1$)} & \makecell{PSNR$\uparrow$\\($\tau=0$)} & \makecell{SSIM$\uparrow$\\($\tau=0$)} & \makecell{LPIPS$\uparrow$\\($\tau=0.9$)} & \makecell{Consistency $\uparrow$\\($\tau=0.9$)} & LR-PSNR$\uparrow$ \\

\hline
\hline
1 & $\mathbf{z}_2,\mathbf{z}_1$ & - & - & 4.76 & 0.34 & 0.863 & 10.56 & - \\ %
\hline
2 & $\mathbf{z}_2,\mathbf{z}_1$  & $\mathbf{y}^*$ & $\mathbf{y}^*,\mathbf{y}_1$ & 28.73 & 0.81 & 0.123 & 41.97 & -\\ %
\hline
3 & $\mathbf{y}_2,\mathbf{z}_2,\mathbf{z}_1$ & $\mathbf{y}^*$ & $\mathbf{y}^*,\mathbf{y}_1$ & 28.71 & 0.81 & 0.124 & 41.79 & 52.77\\ %
\hline
4 & $\mathbf{y}_2,\mathbf{z}_2,\mathbf{z}_1$ & - & - & 18.95 & 0.47 & 0.361 & 40.79 & 53.88 \\ %
\hline
5 & $\mathbf{y}_2,\mathbf{z}_2,\mathbf{z}_1$ & $\mathbf{y}_2$ & $\mathbf{y}_1$ & 28.60 & 0.80 & 0.126 & 41.94 & 52.19\\ %
\hline
\textbf{HCFlow} &  $\mathbf{y}_2,\mathbf{z}_2,\mathbf{z}_1$ & $\mathbf{y}_2$ & $\mathbf{y}_2,\mathbf{y}_1$ & 28.71 & 0.81 & 0.124 & 42.01 & 53.37\\
\hline
\end{tabular}
\end{center}
\end{table*}

\begin{table*}[!thbp]
\captionsetup{font=small}
\footnotesize
\center
\begin{center}
\caption[Caption for LOF]{General image SR ($\times 4$) results on DIV2K~\cite{DIV2K} validation set. For SRFlow and our method, the mean results of 5 draws are reported.}\vspace{-0.2cm}
\label{tab:general_sr}

\begin{tabular}{p{2.7cm}|p{1.cm}<{\centering}|p{1.cm}<{\centering}|p{1.cm}<{\centering}|p{1.cm}<{\centering}|p{1.cm}<{\centering}|p{1.2cm}<{\centering}|p{1.2cm}<{\centering}|p{1.4cm}<{\centering}|p{1.5cm}<{\centering}}
\hline
Method & \#Param & PSNR$\uparrow$ & SSIM$\uparrow$ & LPIPS$\downarrow$ & NIQE$\downarrow$ & BRISQUE$\downarrow$ & Diversity$\uparrow$ & Consistency$\uparrow$ & LR-PSNR$\uparrow$
\\
\hline
\hline
Bicubic  & - & 26.70 & 0.77 & 0.409 & 5.20 & 53.8 & 0 & 38.70 & - \\
EDSR~\cite{lim2017edsr} & 43.1M & 28.98 & 0.83 & 0.270 & 4.46 & 43.3 & 0 & 54.89 & - \\
RRDB~\cite{wang2018esrgan} & 16.7M & 29.44 & 0.84 & 0.253 & 5.08 & 52.4 & 0 & 49.20 &  - \\
ESRGAN~\cite{wang2018esrgan} & 16.7M & 26.22 & 0.75 & 0.124 & 2.61 & 22.7 & 0 & 39.03 & - \\
RankSRGAN~\cite{zhang2019ranksrgan} & 13.7M & 26.55 & 0.75 & 0.128 & 2.45 & 17.2 & 0 & 42.33 &  - \\
SRFlow, $\tau=0$~\cite{lugmayr2020srflow} & 39.5M & 29.07 & 0.81 & 0.254 & 5.20 & 39.4  & 0 & 55.13 & - \\
SRFlow, $\tau=0.9$~\cite{lugmayr2020srflow} & 39.5M & 27.09 & 0.76 & 0.121 & 3.57 & 17.8 & 5.6 & 49.96 & - \\
\hline
\textbf{HCFlow}, $\tau=0$ & 23.2M & 28.71 & 0.81 & 0.285 & 4.61 & 44.1 &  0 & 42.03 & 53.37 \\ %
\textbf{HCFlow}, $\tau=0.9$ & 23.2M & 27.02 & 0.76 & 0.124 & 2.79 & 21.7 & 4.8 & 42.01 & 53.37 \\
\textbf{HCFlow+}, $\tau=0$ & 23.2M & 29.25 & 0.83 & 0.212 & 4.45 & 43.2 & 0 & 51.11 & 53.95 \\
\textbf{HCFlow++}, $\tau=0.9$ & 23.2M & 26.61 & 0.74 & 0.110 & 2.85 & 22.0 & 5.2 & 50.07 & 52.59  \\
\hline
\end{tabular}
\end{center}
\end{table*}

\begin{table*}[!thbp]
\captionsetup{font=small}
\footnotesize
\center
\begin{center}
\caption[Caption for LOF]{Face image SR ($\times 8$) results on CelebA~\cite{liu2015celeba} testing set. For SRFlow and our method, the mean results of 5 draws are reported.}\vspace{-0.2cm}
\label{tab:face_sr}
\begin{tabular}{p{2.6cm}|p{1.0cm}<{\centering}|p{1.cm}<{\centering}|p{1.cm}<{\centering}|p{1.cm}<{\centering}|p{1.cm}<{\centering}|p{1.2cm}<{\centering}|p{1.2cm}<{\centering}|p{1.4cm}<{\centering}|p{1.5cm}<{\centering}}
\hline
Method & \#Param & PSNR$\uparrow$ & SSIM$\uparrow$ & LPIPS$\downarrow$ & NIQE$\downarrow$ & BRISQUE$\downarrow$ & Diversity$\uparrow$ & Consistency$\uparrow$ & LR-PSNR$\uparrow$
\\ 
\hline
\hline
Bicubic  & - & 23.15 & 0.63 & 0.517 & 7.82 & 58.6 & 0 & 35.19 & - \\
RRDB~\cite{wang2018esrgan} & 16.7M & 26.59 & 0.77 & 0.230 & 6.02 & 49.7 & 0 & 48.22  & - \\
ESRGAN~\cite{wang2018esrgan} & 16.7M & 22.88 & 0.63 & 0.120 & 3.46 & 23.7 & 0 & 34.04 & - \\
SRFlow, $\tau=0$~\cite{lugmayr2020srflow} & 40.0M & 26.74 & 0.76 & 0.216 & 5.74 & 40.4 & 0 & 56.57 & - \\
SRFlow, $\tau=0.8$~\cite{lugmayr2020srflow} & 40.0M & 25.24 & 0.71 & 0.110 & 4.20 & 23.2 & 5.2 & 50.85 & - \\
\hline
\textbf{HCFlow}, $\tau=0$ & 27.0M & 26.66 & 0.77 & 0.210 & 6.42 & 48.0 & 0 & 51.83 & 54.50 \\ %
\textbf{HCFlow}, $\tau=0.8$ & 27.0M & 24.99 & 0.71 & 0.104 & 4.34 & 31.6 & 5.9 & 51.81 & 54.50 \\
\textbf{HCFlow+}, $\tau=0$ & 27.0M & 27.02 & 0.78 & 0.212 & 6.04 & 49.5 & 0 & 51.11 & 53.95\\
\textbf{HCFlow++}, $\tau=0.8$ & 27.0M & 24.83 & 0.69 & 0.090 & 3.87 & 23.8  & 4.0 & 51.57 & 51.82 \\
\hline
\end{tabular}
\end{center}
\end{table*}

\vspace{-0.4cm}
\paragraph{Image rescaling.}
For image rescaling ($\times 4$), we set $L, K, P$ to 2, 8 and 6, respectively. Two 3-block RRDB networks are used as feature extractors. In particular, we use Haar transformation to replace the squeeze layer and remove invertible $1\times 1$ convolution layers. Details on data preparation and optimizer are the same as general image SR. The learning rate is initialized as $2.5\times 10^{-4}$ and halved at $[100k,200k,300k, 400k]$ ($500k$ iterations in total). The loss weighting parameters are $\lambda_1=1$, $\lambda_2=5\times 10^{-2}$ and $\lambda_3=1\times 10^{-5}$, respectively.

\vspace{-0.4cm}
\paragraph{Performance evaluation.} 
Following SRFlow~\cite{lugmayr2020srflow} and IRN~\cite{xiao2020IRN}, we evaluate PSNR and SSIM on the RGB color space for image SR, and on the Y channel of the YCbCr color space for image rescaling. We also use perceptual metric LPIPS~\cite{zhang2018lpips} and two no-reference metrics, NIQE~\cite{mittal2012NIQE} and BRISQUE~\cite{mittal2011BRISQUE}, for better visual quality comparison. Pixel standard deviation of 5 samples are used to compare the diversity of results. In addition, Consistency (PSNR between the downscaled SR image and the ground-truth LR image) and LR-PSNR (PSNR between the generated LR image in forward propagation and the ground-truth LR image) are also reported.

\subsection{Ablation Study}

\vspace{-0.1cm}
\paragraph{Fitting to the LR image space.}
To learn a fully invertible flow between HR and LR image pairs, HCFlow constrains part of the latent space to be the LR image space, instead of using the LR image as an external prior. To show the impact, we remove the LR image $\textbf{y}_2$ from the latent space as shown in case 1 and 2 of Table~\ref{tab:ablation_latent}. When there is no conditional prior (case 1), the model fails to converge as it does not have enough information for SR . When we replace $\textbf{y}_2$ with ground-truth LR image $\textbf{y}^*$ as a conditional prior (case 2, similar to SRFlow~\cite{lugmayr2020srflow}), it achieves slightly better performance than HCFlow although they have almost the same conditional information. The underlying reason might be that it has a larger latent space than HCFlow.

\vspace{-0.4cm}
\paragraph{Ground-truth LR image as a conditional prior.}
HCFlow is conditional on $\mathbf{y}_1$ and $\mathbf{y}_2$, which are generated during propagation. When we use the ground-truth LR image $\mathbf{y}^*$ as a conditional prior to replace $\mathbf{y}_2$ (case 3, Table~\ref{tab:ablation_latent}), the model achieves similar performance as HCFlow. In fact, since we model the distribution of $\mathbf{y}_2$ as a Dirac delta function $\delta(\mathbf{y}_2-\mathbf{y}^*)$, $\mathbf{y}_2$ would be nearly equal to $\mathbf{y}^*$ after model convergence, which is confirmed by the high LR-PSNR. Therefore, conditional on the generated $\mathbf{y}_2$ and the external $\mathbf{y}^*$ have similar effects.

\begin{figure*}[!htbp]
\captionsetup{font=small}
\scriptsize
\hspace{-0.2cm}
\begin{tabular}{c@{\extracolsep{0em}}@{\extracolsep{0.05em}}c@{\extracolsep{0.05em}}c@{\extracolsep{0.05em}}c@{\extracolsep{0.05em}}c@{\extracolsep{0.00em}}c@{\extracolsep{0.00em}}|@{\extracolsep{0.25em}}c}
		\includegraphics[width=0.137\textwidth]{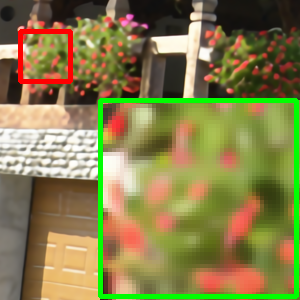}~
		&\includegraphics[width=0.137\textwidth]{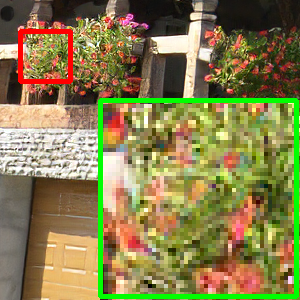}~
		&\includegraphics[width=0.137\textwidth]{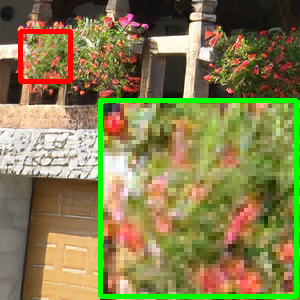}~
		&\includegraphics[width=0.137\textwidth]{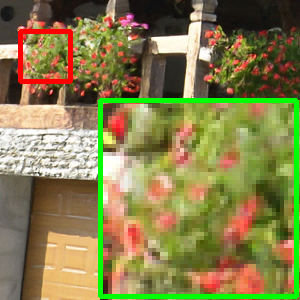}~
		&\includegraphics[width=0.137\textwidth]{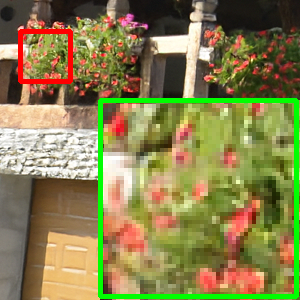}~
		&\includegraphics[width=0.137\textwidth]{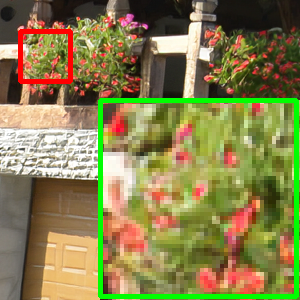}~
		&\includegraphics[width=0.137\textwidth]{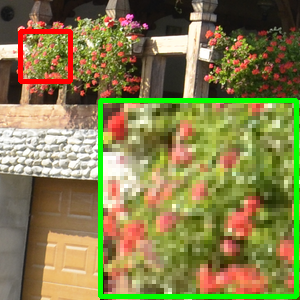}~\\

		\includegraphics[width=0.137\textwidth]{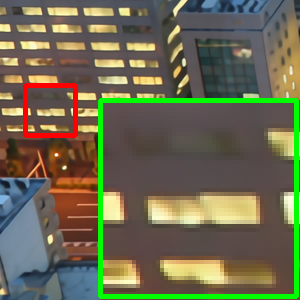}~
		&\includegraphics[width=0.137\textwidth]{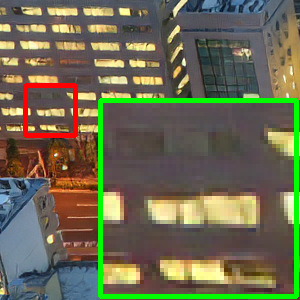}~
		&\includegraphics[width=0.137\textwidth]{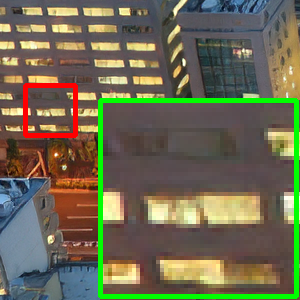}~
		&\includegraphics[width=0.137\textwidth]{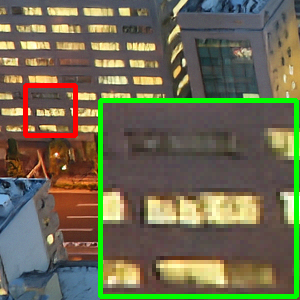}~
		&\includegraphics[width=0.137\textwidth]{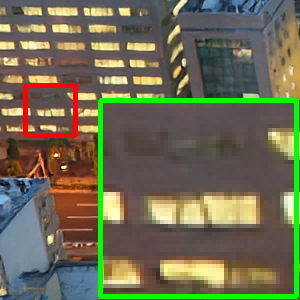}~
		&\includegraphics[width=0.137\textwidth]{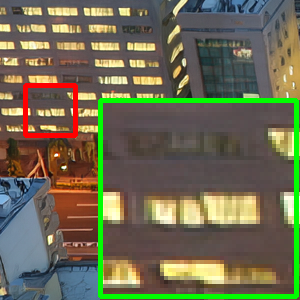}~
		&\includegraphics[width=0.137\textwidth]{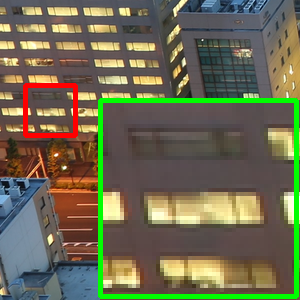}~\\
		
		\includegraphics[width=0.137\textwidth]{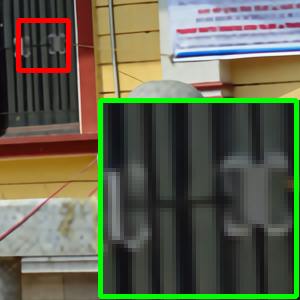}~
		&\includegraphics[width=0.137\textwidth]{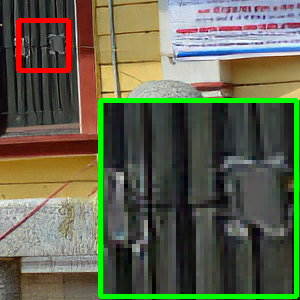}~
		&\includegraphics[width=0.137\textwidth]{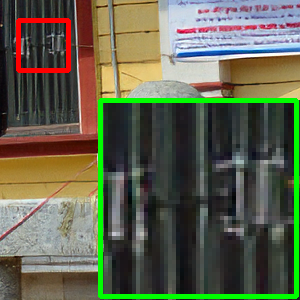}~
		&\includegraphics[width=0.137\textwidth]{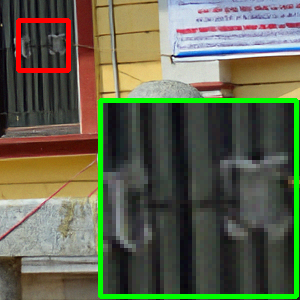}~
		&\includegraphics[width=0.137\textwidth]{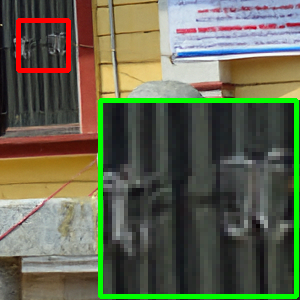}~
		&\includegraphics[width=0.137\textwidth]{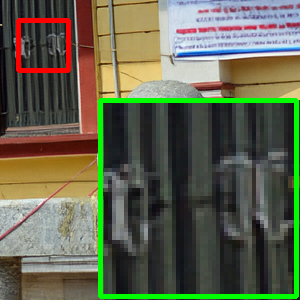}~
		&\includegraphics[width=0.137\textwidth]{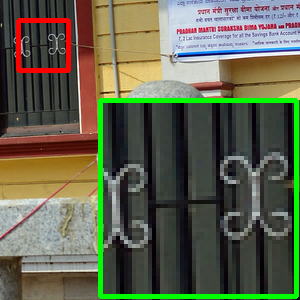}~\\
 RRDB & ESRGAN & RankSRGAN & SRFlow, $\tau=0.9$  & \makecell{\textbf{HCFlow}, $\tau=0.9$} & \makecell{\textbf{HCFlow++}, $\tau=0.9$}  & Ground Truth \\
\end{tabular}
\vspace{-0.2cm}
\caption{Visual results of general image SR ($\times 4$) on the DIV2K~\cite{DIV2K} validation set.}
\label{fig:general_visualresults}
\end{figure*}

\begin{figure*}[!htbp]
\captionsetup{font=small}
\scriptsize
\hspace{-0.20cm}
\begin{tabular}{c@{\extracolsep{0em}}@{\extracolsep{0.05em}}c@{\extracolsep{0.05em}}c@{\extracolsep{0.05em}}c@{\extracolsep{0.05em}}c@{\extracolsep{0.00em}}c@{\extracolsep{0.00em}}|@{\extracolsep{0.25em}}c}
		
		\includegraphics[width=0.137\textwidth]{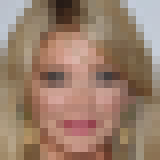}~
		&\includegraphics[width=0.137\textwidth]{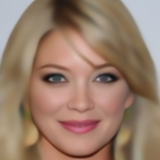}~
		&\includegraphics[width=0.137\textwidth]{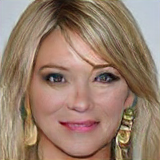}~
		&\includegraphics[width=0.137\textwidth]{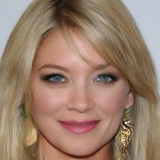}~
		&\includegraphics[width=0.137\textwidth]{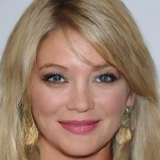}~
		&\includegraphics[width=0.137\textwidth]{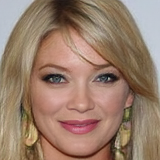}~
		&\includegraphics[width=0.137\textwidth]{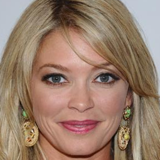}~\\

		\includegraphics[width=0.137\textwidth]{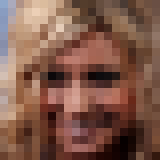}~
		&\includegraphics[width=0.137\textwidth]{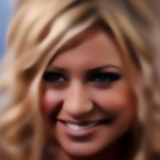}~
		&\includegraphics[width=0.137\textwidth]{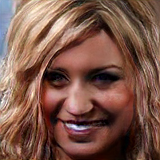}~
		&\includegraphics[width=0.137\textwidth]{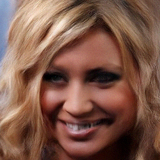}~
		&\includegraphics[width=0.137\textwidth]{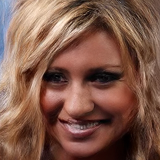}~
		&\includegraphics[width=0.137\textwidth]{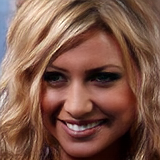}~
		&\includegraphics[width=0.137\textwidth]{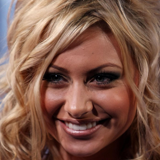}~\\
		
		\includegraphics[width=0.137\textwidth]{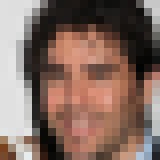}~
		&\includegraphics[width=0.137\textwidth]{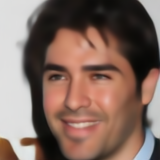}~
		&\includegraphics[width=0.137\textwidth]{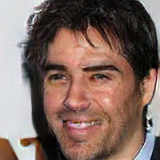}~
		&\includegraphics[width=0.137\textwidth]{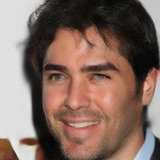}~
		&\includegraphics[width=0.137\textwidth]{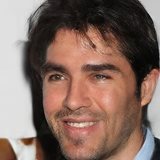}~
		&\includegraphics[width=0.137\textwidth]{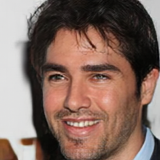}~
		&\includegraphics[width=0.137\textwidth]{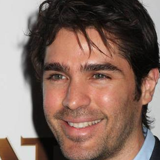}~\\

LR ($\times$8) & RRDB & ESRGAN & SRFlow, $\tau=0.8$  & \makecell{\textbf{HCFlow}, $\tau=0.8$} & \makecell{\textbf{HCFlow++}, $\tau=0.8$} & Ground Truth \\
\end{tabular}
\vspace{-0.2cm}
\caption{Visual results of face image SR ($\times 8$) on the CelebA~\cite{liu2015celeba} testing set.}
\label{fig:face_visualresults}
\vspace{-0.2cm}
\end{figure*}

\vspace{-0.4cm}
\paragraph{Hierarchical conditional mechanism.}
As shown in case 4 of Table~\ref{tab:ablation_latent}, similar to IRN~\cite{xiao2020IRN}, we assume the LR image and the rest high-frequency component is independent by removing all conditional priors. It yields significantly worse performance because the reconstruction of HR image (high-frequency component) is highly conditional on the LR image (low-frequency component) for image SR. Despite this, it has better results than case 1, as fitting to the LR image space could partly play the role of conditional prior. In case 5, we change from hierarchical conditional mechanism to single-scale conditional mechanism, by removing $\mathbf{y}_2$ from level 1. In this case, $\mathbf{z}_l$ ($l=1,2$) is only conditional on $\mathbf{y}_l$ from the same level. The performance drops in terms of all kinds of metrics, which shows that the hierarchical conditional mechanism can better model the conditional relations between high-frequency and low-frequency components.

\subsection{Experiments on Image SR}
\vspace{-0.1cm}
\paragraph{General image SR.}
For general image SR ($\times 4$), we compare HCFlow with state-of-the-art CNN-based and flow-based SR models, including the PSNR-oriented EDSR~\cite{lim2017edsr} and RRDB~\cite{wang2018esrgan}, perception-oriented ESRGAN~\cite{wang2018esrgan} and RankSRGAN~\cite{zhang2019ranksrgan}, as well as SRFlow~\cite{lugmayr2020srflow}. All methods are trained on the same training dataset. From Table~\ref{tab:general_sr} and Fig.~\ref{fig:general_visualresults}, we have several observations as follows. First, when sampling HR images with temperature $\tau=0$, HCFlow acts like a PSNR-oriented model, achieving similar performance as EDSR and RRDB. Adding the HR pixel loss (\ie, HCFlow+) can further improve the PSNR and SSIM by large margins. Second, when $\tau=0.9$, the perceptual metrics of HCFlow are boosted dramatically. With perceptual loss and GAN loss (\ie, HCFlow++), the perceptual metrics are further improved by significant margins in terms of LPIPS and BRISQUE, which is confirmed by the visual results. Note that, unlike ESRGAN and RankSRGAN, the generated HR images of HCFlow++ are still diversified. Third, HCFlow achieves state-of-the-art performance in terms of both quantitative metrics and visual quality. It generates sharp images with few artifacts. In contrast, RRDB and SRFlow tend to produce blurry images, while ESRGAN and RankSRGAN suffer from over-sharpen artifacts and distortions. In addition, HCFlow only has about half of the number of parameters compared with SRFlow.

\begin{table*}[!thbp]
\captionsetup{font=small}
\footnotesize
\center
\begin{center}
\caption[Caption for LOF]{Image rescaling ($\times 4$) results (Y-channel PSNR / SSIM) on different datasets. For IRN~\cite{xiao2020IRN} and our method, the mean results of 5 draws are reported. Differences of PSNR / SSIM of different samples are less than 0.02.}\vspace{-0.2cm}
\label{tab:ir_psnr}
\begin{tabular}{p{3.2cm}|p{1.1cm}<{\centering}|p{2cm}<{\centering}|p{2cm}<{\centering}|p{2cm}<{\centering}|p{2cm}<{\centering}|p{2cm}<{\centering}}
\hline
Downscaling \& Upscaling & Param & Set5~\cite{Set5} &  Set14~\cite{Set14} &  BSD100~\cite{BSD100} &  Urban100~\cite{Urban100} & DIV2K~\cite{DIV2K}
\\
\hline
\hline
Bicubic \& Bicubic & - & 28.42 / 0.8104 &26.00 / 0.7027 &25.96 / 0.6675 &23.14 / 0.6577 &26.66 / 0.8521 \\
Bicubic \& SRCNN~\cite{dong2014srcnn} & 57.3K &30.48 / 0.8628 &27.50 / 0.7513 &26.90 / 0.7101 &24.52 / 0.7221 &– \\
Bicubic \& RDN~\cite{zhang2018RDN} & 22.3M &32.47 / 0.8990 &28.81 / 0.7871 &27.72 / 0.7419 &26.61 / 0.8028 &–\\
Bicubic \& EDSR~\cite{lim2017edsr} & 43.1M &32.62 / 0.8984 &28.94 / 0.7901 &27.79 / 0.7437 &26.86 / 0.8080 &29.38 / 0.9032\\
Bicubic \& RCAN~\cite{zhang2018rcan} & 15.6M &32.63 / 0.9002 &28.87 / 0.7889 &27.77 / 0.7436 &26.82 / 0.8087 &30.77 / 0.8460\\
Bicubic \& RFANet~\cite{liu2020RFANet} & 11.2M &32.67 / 0.9004 &28.88 / 0.7894 &27.79 / 0.7442 &26.92 / 0.8112 &–\\
Bicubic \& RRDB~\cite{wang2018esrgan} & 16.3M &32.74 / 0.9012 &29.00 / 0.7915 &27.84 / 0.7455 &27.03 / 0.8152 &30.92 / 0.8486\\
TAD \& TAU~\cite{kim2018task} & – &31.81 / – &28.63 / – &28.51 / – &26.63 / – &31.16 / –\\
CAR \& EDSR~\cite{sun2020learned} & 52.8M &33.88 / 0.9174 &30.31 / 0.8382 &29.15 / 0.8001 &29.28 / 0.8711 &32.82 / 0.8837\\
IRN~\cite{xiao2020IRN} & 4.4M & {36.19 / 0.9451} & {32.67 / 0.9015} & {31.64 / 0.8826} & {31.41 / 0.9157} & {35.07 / 0.9318}\\
\hline
\textbf{HCFlow} & 4.4M & {36.29 / 0.9468} & {33.02 / 0.9065} & {31.74 / 0.8864} & {31.62 / 0.9206} & {35.23 / 0.9346} \\ %

\hline
\end{tabular}
\end{center}
\vspace{-0.2cm}
\end{table*}

\vspace{-0.5cm}
\paragraph{Face image SR.}
We also test HCFlow on face image SR ($\times 8$) to show its effectiveness. The compared methods include PSNR-oriented RRDB, perception-oriented ESRGAN and the flow-based SRFlow. As shown in Table~\ref{tab:face_sr} and Fig.~\ref{fig:face_visualresults}, similar observations as in general image SR can be concluded for face image SR. HCFlow achieves best quantitative and visual performance compared with competing methods. In particular, HCFlow generates sharp faces with natural details, especially on eyes, teeth and hairs. By comparison, other methods suffer from either over-smoothed results or obvious artifacts.

\subsection{Experiments on Image Rescaling}
As a unified framework for image SR and image rescaling, HCFlow also achieves state-of-the-art performance in image rescaling. We compare it with three kinds of rescaling methods: (1) bicubic interpolation \& state-of-the-art SR models~\cite{dong2014srcnn, zhang2018RDN, lim2017edsr, zhang2018rcan, wang2018esrgan, liu2020RFANet}; (2) encoder-decoder models~\cite{kim2018task, sun2020learned}; (3) invertible neural networks~\cite{xiao2020IRN}. 

\begin{figure}[!t]
\captionsetup{font=small}
\scriptsize
	\vspace{0.2cm}
 \hspace{-0.24cm}
\begin{tabular}{c@{\extracolsep{0em}}@{\extracolsep{0.05em}}c@{\extracolsep{0.05em}}c@{\extracolsep{0.05em}}c@{\extracolsep{0.05em}}c@{\extracolsep{0.00em}}c@{\extracolsep{0.00em}}|@{\extracolsep{0.25em}}c}
		\includegraphics[width=0.155\textwidth]{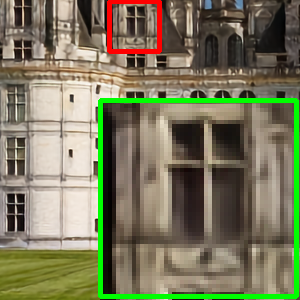}~
		&\includegraphics[width=0.155\textwidth]{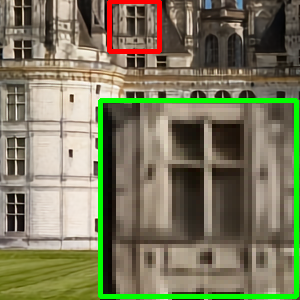}~
		&\includegraphics[width=0.155\textwidth]{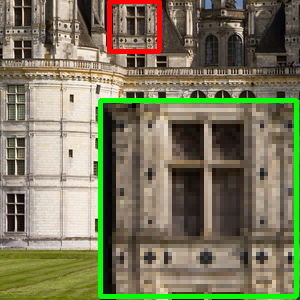}~\\

		\includegraphics[width=0.155\textwidth]{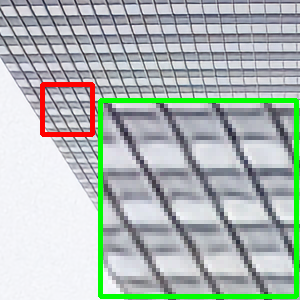}~
		&\includegraphics[width=0.155\textwidth]{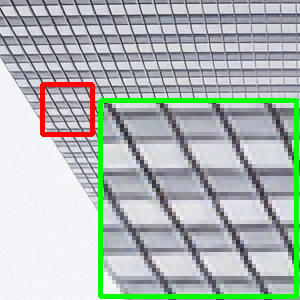}~
		&\includegraphics[width=0.155\textwidth]{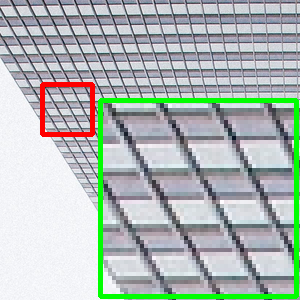}~\\
		
		\includegraphics[width=0.155\textwidth]{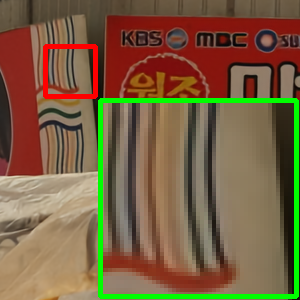}~
		&\includegraphics[width=0.155\textwidth]{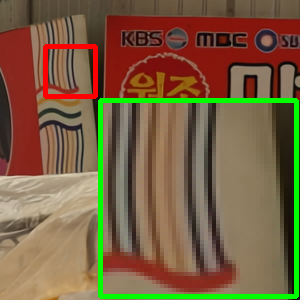}~
		&\includegraphics[width=0.155\textwidth]{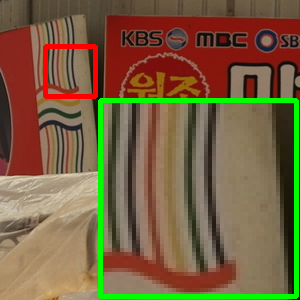}~\\
		
IRN & HCFlow & Ground Truth \\
\end{tabular}
\vspace{-0.2cm}
\caption{Visual results of image rescaling ($\times 4$) on the DIV2K~\cite{DIV2K} validation set. More results are shown in the supplementary.}
\label{fig:ir_visualresults}
\vspace{-0.5cm}
\end{figure}

As can be seen from Table~\ref{tab:ir_psnr}, when the downscaling process is fixed (\ie, bicubic interpolation), performances of different state-of-the-art SR models are similar and limited. When the downscaling models are optimized for the upscaling models, the results are largely improved. IRN further boosts the performance by joint optimization based on the invertible architecture. Compared with IRN, the proposed HCFlow achieves better performance on all testing datasets with an increased PSNR of $0.10\sim0.35dB$. Besides, as shown in Fig.~\ref{fig:ir_visualresults}, HCFlow can better preserve image details and generates sharper edges than IRN. Since these two models have same number of parameters, HCFlow is more efficient than IRN for image rescaling, which can be attributed to the conditional modelling between high-frequency and low-frequency components.

\section{Conclusion}
In this paper, we proposed a unified framework, \ie, hierarchical conditional flow (HCFlow), for both image super-resolution and image rescaling. It learns a fully invertible mapping between HR image and LR image as well as the latent variable. Particularly, we learn the LR image space and design a hierarchical conditional mechanism between the latent variable (high-frequency component) and the LR image (low-frequency component). For image SR, HCFLow is trained by the negative log-likelihood loss, and is further enhanced by pixel loss, perceptual loss and GAN losses for better performance. For image rescaling, it is trained as an encoder-decoder framework, where the forward and inverse progresses are jointly optimized. Experiments demonstrate that HCFlow achieves state-of-the-art performance on general image SR, face image SR and image rescaling, in terms of both quantitative metrics and visual quality.

\vspace{0.2cm}
\noindent\textbf{Acknowledgements}~~ We thank Dr. Suryansh Kumar for helpful discussion. This work was partially supported by the ETH Zurich Fund (OK), a Huawei Technologies Oy (Finland) project, the China Scholarship Council and a Microsoft Azure grant. Special thanks goes to Yijue Chen.

{\small
\bibliographystyle{ieee_fullname}
\bibliography{superresolution.bib}
}

\end{document}